# Polymer conformation: genetic and environmental factors associated with this characteristic of glutenin in wheat grain

Gérard Branlard [1*], Angelina d'Orlando [2], Ayesha Tahir [3], Marc Schmutz [4], Larbi Rhazi [5], Annie Faye [1] and Thierry Aussenac [5*]

1 The French National Research Institute for Agriculture, Food and the Environment (INRAE), UCA UMR1095 GDEC, 5 Chemin de Beaulieu, 63100 Clermont-Ferrand, France; gerard.branlard@gmail.com (GB); annie.faye@inrae.fr (AF)

2 The French National Research Institute for Agriculture, Food and the Environment (INRAE), Unité BIA-Plateforme BIBS, 3 Impasse Yvette Cauchois, 44 316 Nantes, France
angelina.dorlando@inrae.fr (AD)

3 Department of Biosciences, COMSATS University Islamabad, Park Road, Tarlai Kalan, 45550 Islamabad, Pakistan; ayesha.tahir2007@gmail.com (AT)

4 Université de Strasbourg, CNRS, Institut Charles Sadron, 23 rue du Loess, B.P. 84047, 67034 Strasbourg Cedex, France marc.schmutz@ics-cnrs.unistra.fr (MS)

5 Institut Polytechnique UniLaSalle, Université d'Artois, ULR 7519, 19 rue Pierre Waguet, BP 30313, 60026 Beauvais, France; larbi.rhazi@unilasalle.fr (LR), thierry.aussenac@unilasalle.fr (TA)

*Correspondence:     gerard.branlard@gmail.com

thierry.aussenac@unilasalle.fr;

**Highlights**

Original photos of the glutenin polymers (GPs) are shown.

Thousands of billions GPs are synthetized in each wheat grain.

GPs become bigger, more compact and less reticulated with higher temperature.

**Abstract**

Using asymmetric flow field flow fractionation, the polymer mass (Mw), gyration radius (Rw) and the polydispersity index (PI) of glutenin polymers (GPs) have been previously studied. Here, using the same multi-location trials (4 years, 11 locations and 192 cultivars), the factors associated with the conformation (Conf) of the polymers were reported. Conf, which is the slope of Log (Rw) = f [Log (Mw)], varied between 0.285 and 0.740. Conf had low broad-sense heritability ($H^2$ = 16.8) and was significantly influenced by temperature occurring over the last month of grain filling. Higher temperatures were found to increase Rw and the compactness and sphericity of GPs. The HMW-GS and LMW-GS alleles had a significant influence on the Conf value. Assuming a Gaussian distribution for Mw, the number of polymers present in wheat grain was computed for different kernel weights and protein concentrations: it exceeded $10^{12}$ GPs per grain. Using atomic force microscopy and cryo-microscopy, photos of GPs were revealed for the first time. Under high temperatures, GPs became larger, less reticulated, more spherical and consequently less prone to rapid hydrolysis. Some orientations aimed at reducing the possible impact of the numerous GPs for people suffering from non-celiac gluten sensitivity (NCGS) were proposed.

**Introduction**

Wheat dough has been used for more than two millenniums to make leavened bread. Due to the unique properties of the gluten network formed during dough mixing, it can retain gas. Grain storage proteins (GSPs), *i.e.*, monomeric gliadins and polymeric glutenins, are well known to confer viscosity and elasticity, respectively, to wheat gluten (Finney, 1943); these properties are known to be largely influenced by wheat-growing conditions. The effect of nitrogen uptake and, more recently, the temperature occurring during GSP accumulation have been studied using both transcriptomics tools (Perrotta *et al.*, 1998; Hurkman *et al.*, 2003; Altenbach *et al.*, 2007) and proteomics approaches (Skylas *et al.*, 2002; Majoul *et al.*, 2003, 2004, 2013; Vensel *et al.*, 2005; Hurkman *et al.*, 2009; Laino *et al.*, 2010). High temperatures shorten the effective grain filling period and may significantly increase the protein concentration because of the more dramatic effect on starch than on protein accumulation (Blumenthal *et al.*, 1991a, 1991b; Corbellini *et al.*, 1998; Altenbach *et al.*, 2003; Triboï *et al.*, 2003). For example, in response to heat shock (38°C for 4 h on for 4 consecutive days during grain filling), the volume occupied by the A starch granules compared to the normal control increased from 71 to 78.9%, whereas the B and C granules decreased from 27.8 to 20.1% and 1.2 to 1.0%, respectively, in mature grain (Branlard *et al.*, 2015). The synthesis of individual wheat GSPs, gliadins and glutenin, was not drastically affected by high temperature as opposed to glutenin polymers, analysed for the first time in only 6 of our 11 locations (Branlard *et al.*, 2015).

Since the 1970s, numerous genetic and molecular GSP studies have been devoted to helping breeders better respond to the demand raised by technologists for flour use. Many advances in GSPs, which are the major gluten components and their influence on wheat quality, have recently been gathered in an international collaborative book (Igrejas *et al.*, 2020). Moreover, a better understanding of the developing endosperm, particularly in stressed conditions resulting from its hypoxic status, came from studies carried out in maize (Young and Gallie, 2000; Rolletschek *et al.*, 2005; Gayral *et al.*, 2017), rice (Xu *et al.*, 2008) and wheat (Zhen *et al.*, 2016). However, the technological abilities of wheat grain to make bread, biscuit or cookies for *T. aestivum* cultivars and pasta and semolina for *T. durum* cultivars are only a part of the final wheat quality, which must respond to consumer needs.

Numerous approaches have been carried out to improve the nutritional and health value of wheat. These studies include the improvement of the lysine content, an essential amino acid (Shewry and Hey, 2015), mineral and vitamin content mostly in the aleurone layer (Meziani *et al.*, 2021), and healthy arabinoxylan and oligosaccharides content and the removal or reduction of numerous allergenic compounds, such as alpha amylase / trypsin inhibitors and the numerous epitopes in proteins involved in specific pathologies, including celiac disease (prevalence 1%) and baker asthma (first disease associated with a professional activity in Europe). These important requirements, which have received recent updates (Brouns *et al.*, 2019), will not be developed in the present study, focusing on gluten polymer conformation. Non-celiac gluten-related disorders, called non-celiac gluten sensitivity (NCGS), are a public concern (Catissi *et al.*, 2013). Numerous NCGS symptoms are expressed by an increasing number of people, and their prevalence has increased, primarily attributable to gluten.

No specific sequences of GSP—*i.e.*, either from gliadins or from glutenin subunits of high or low molecular weight (HMW-GS and LMW-GS, respectively), which could be characteristics of immune response—were significantly associated with NCGS (Scherf *et al.*, 2016). Consequently, several alternative hypotheses have been investigated to explain NCGSs (Catissi *et al.*, 2013; Scherf *et al.*, 2016), such as the $\alpha$-amylase and trypsin inhibitors (ATIs), and fermentable oligo, di-, and monosaccharides and polyols (FODMAPS) (Catassi, 2015; for review, see Brouns *et al.*, 2019). The re-orientation of these hypotheses was probably due to the absence of dedicated tools used for glutenin polymer characterisation. When wheat varieties are studied for human health, the amount of insoluble polymers or unextractable polymeric protein (%UPP) is usually obtained after a two-step extraction using SDS, followed by sonication through SE-HPLC (Gupta *et al.*, 1993, Johansson *et al.*, 2013). However, the maximum size sieving offered by chromatography columns is around $1 \times 10^6$ Da. The use of Asymmetric Flow Field Flow Fractionation Multi-Angular Laser Light Scattering (A4F-MALLS) allows for a deeper analysis of the polymer characteristics whose mass exceeds $10^7$ Da (Wahlund *et al.*, 1996; Lemelin *et al.*, 2005). A first attempt to measure the conformation of wheat polymers was performed using size exclusion chromatography (SEC-MALLS), revealing a conformation close to the theoretical compact sphere. Analytical difficulties (pH and SEC columns) have prevented the use of SEC-MALLS in routine analysis (Mendichi *et al.*, 2008). A4F provides several characteristics of chemical macromolecules that can be very useful for understanding their technological properties, as well as their functional properties. Using A4F, the polymer mass (Mw), gyration radius (Rw), and polydispersity index (PI) of polymers were measured over a four-year, multi-location wheat trial in France, involving 11 locations and 192 cultivars, which revealed that Rw (MW and PI) increased linearly (exponentially) with the average mean temperature, registered hourly, of the last month of the wheat cycle (Branlard *et al.*, 2020). The highly significant increases of 189, 242 and 434% of the Mw, Rw and PI, respectively, were caused by a 3.5°C increase between trial locations. These findings attract more consideration on glutenin polymer characteristics. To this end, an update on the structure and assembly of glutenin polymers (GPs) has recently been proposed (Shewry and Lafiandra, 2022). Variations in GP characteristics, particularly regarding the response to global warming, require investigation. In addition to these 3 characteristics, little is known about the shape of glutenin polymers. What are the environmental and genetic influences on the conformation of the GPs; are they compact, partly reticulated or highly depolymerised with a fibrillar shape? Multi-angular laser light scattering computes the conformation (Conf) values of the polymers. Classic analyses of chemical polymers usually report the following classification of the polymer shape as a function of the Conf values, also called slope. For example, a coefficient < 0.30 indicates a compact spherical shape; a slope value between 0.5 and 0.6 corresponds to a polymer with random coils (Wyatt, 1993). These characteristics corresponding to the physical A4F measures of the polymers were also compared to the visual observation using two specific techniques: atomic force microscopy (AFM) previously used in $H_2O/D_2O$ medium (Haward *et al.*, 2011) and cryo-transmission electron microscopy (Cryo-TEM). These studies revealed, for the first time, glutenin polymers in lyophilised conditions and in solution, respectively. In addition, based on the masses of polymers measured by A4F and assuming their Gaussian distribution in wheat samples, the number of polymers was computed for wheat grains with different kernel weights and protein concentrations.

The polymer conformation (Conf) is the fourth parameter of the polymers, which was analysed using A4F on the same multi-location samples, allowing us to test factors associated with phenotypic variations of Conf. The influences of genetic factors (the 14 and 17 alleles encoding HMW-GS and LMW-GS, respectively) and environmental factors, such as rainfall and temperature, registered daily during the last three months of the wheat cycle were studied. The values of the Conf parameter may impact the degradation and hydrolysis of glutenin polymers. Consequently, the identification of glutenin alleles associated with the conformation of polymers is of prime importance and will be particularly discussed regarding possible further studies dealing with consumer health.

**Materials and Methods**

*Multi-location trials*

The material used in the present study was described in previous multi-location trials carried out for four years in France between 2004 and 2010 involving 11 locations and 192 hexaploid wheat cultivars originating from France, Germany, the Netherlands and the UK. (Branlard *et al.*, 2020). A total of 885 samples were analysed for several physical and technological characteristics [1000-kernel weight (TKW), test weight (TW), protein concentration (PC) and grain hardness (GH)]. A total of 14 alleles of HMW-GS encoded at the three *Glu-1* loci and 17 alleles of LMW-GS encoded at the three *Glu-3* loci were identified using SDS-Page. The 11 locations of the wheat trials and the frequencies of the 31 alleles are reported in Supplemental Tables S1 - S2, respectively (Branlard *et al.*, 2020). The cumulated daily temperature per month and daily water precipitation per month for May, June and July of the four experimental years are also given in Figures S1 - S2 of the previous publication.

*Determination of the polymer's conformation*

The glutenin polymers were extracted from whole meal flour and separated using the A4F-MALLS technique, as previously described for multi-location trials (Branlard et al., 2020). The following three parameters of the glutenin polymers were determined:

$Mw$ = the average weight-average molar mass, expressed in g mol$^{-1}$ or Da;

$Rw$ = the weight-average mean square radius or hydrodynamic radius, expressed in nm;

$PI = Mw/Mn$ the polydispersity index, where $Mn$ is the average molar mass.

The Conformation characteristic (Conf) is computed by the A4F-MALLS software on the bases described by Wyatt, 1993. Each analysed sample may possess N polymers and the apparatus measures n times the characteristics (mass Mw, radius Rw, etc…) of each polymer. The software compute several additional characteristics including the conformation of the polymer. Then having analysed the N polymers present in a given sample, the Conf value of the polymers in the sample is determined from the slope (1/a) of the following Log-Log plot: abscissa Log (Mw) and in ordinate Log of root-mean square (RMS) of Rw, *ie* : $(\sum_{i=1}^{n} \frac{(Rwi)^2}{n})^{0.5}$. An example of that Log-Log plot for a glutenin sample is shown in Supplemental Figure S1. The characteristics of the chemical polymers according to the Conf values were classically described as follows (Wyatt, 1993):

Conf < 0.33: Compact and spherical; Conf 0.33–0.50: Partly depolymerised and reticulated; Conf > 0.50: Depolymerised and fibrillar; Conf = 1: Polymer shape of a rigid rod.

*Sequential protein extraction for AFM and Cryo-TEM*

The procedure used here was performed according to Fu and Sapirstein (1996) and Rhazi *et al*. (2003). Flour samples (0.125 g) were stirred for 1 h at room temperature with 2.25 ml of 0.08 M Tris–HCl buffer (pH 7.5) containing 50% (v/v) 1-propanol. Samples were centrifuged at 15,900 × g for 30 min at 15°C. Soluble polymeric proteins in 50% 1-propanol were precipitated by the addition of 1.5 ml of 1-propanol to bring the 1-propanol proportion of the supernatant to 70% (v/v). The dispersion was agitated and left to stand at room temperature for 1 h after centrifugation at 26,000 × g for 15 min at 15°C. The supernatant, mainly containing monomeric proteins, was removed. The pellet containing mainly aggregates of polymers was mixed with 1 ml of sodium phosphate buffer (0.1 M, pH 6.9) containing 1.5% (w/v) SDS, sonicated for 20 s at a power setting of 30% with a 3 mm microtip, and then allowed to stand at 60°C for 2 h. After centrifugation (12,500 × g for 30 min at 25°C), the supernatants were used for Cryo-TEM after dilution and freeze-dried for AFM observations.

*Proteomics*

Proteomics analyses carried out on 21 stages of wheat endosperm development of the cultivar 'Récital' grown in a greenhouse under controlled conditions were used to determine the expression of the major enzymes involved in the control of redox and reactive oxygen species (ROS) (Tasleem-Tahir *et al.*, 2012).

*Atomic force microscopy (AFM) of the glutenin polymers*

AFM images (height and error-signal phase images) of glutenin polymers were recorded in air, employing INNOVA AFM from Bruker (Berlin, Germany). The lyophilised polymers obtained from sequential protein extraction were deposited on new silicon substrates or piranha solution-cleaned silicon substrates, whose cleanliness was checked beforehand with optics and AFM. Images were registered with low scan rates (0.3 Hz) using nitride-coated silicon probes with spring constants of 2.8 $N.m^{-1}$ (FESPA-V2, Bruker, Germany). The sizes of the images were all between 5 µm × 5 µm and 10 µm × 10 µm at a resolution of 512 × 512 pixels. Treatment to improve AFM images and height profiles was achieved using Gwyddion Software (Brno, Czech Republic).

*Cryo-transmission electron microscopy (Cryo-TEM) of the glutenin polymers*

The principle of Cryo-TEM of vitrified specimens was previously described (Dubochet *et al.*, 1988). The vitrification of the samples was carried out in a homemade vitrification system. The chamber was maintained at 22°C, and the relative humidity was 80%. A 5 µl drop of the sample was deposited onto a lacey carbon film covered with a 300 mesh Cu grid (Ted Pella-USA) rendered hydrophilic using an ELMO glow discharge unit (Cordouan Technologies, Pessac, France). The grid was automatically blotted to form a thin film and plunged in liquid ethane at −190°C, as maintained by liquid nitrogen. Thus, a vitrified film was obtained in which the native structure of the vesicles was preserved. The grid was mounted onto a cryo holder (Gatan 626-Pleasanton, CA, USA) and observed under low-dose conditions (10 $e/A^2$) in a Tecnai G2 microscope (FEI, Eindhoven, Netherlands) at 200 kV. Images were acquired using an Eagle slow scan CCD camera (FEI).

*Estimation of the number of polymers in wheat grain*

Preliminary considerations:
- The average polymer masses (Mw) measured in wheat samples using A4F were usually between $1 \times 10^6$ and $2 \times 10^8$ Da (or g $mol^{-1}$).
- Wheat storage protein analyses showed that, in general, for bread wheat cultivars, polymerised glutenin with a molecular weight greater that $10^6$ g $mol^{-1}$, varied between 20 and 30% of total glutenins, as usually found in our analytical conditions using A4F (Carceller and Aussenac, 2001; Aussenac *et al.*, 2020). To reduce the number of calculus, we chose the average of 25% of the total quantity of glutenins as part of the polymerised glutenins.
- If all glutenin polymers in the grain had the same identical mass (Mw; for example, $15 \times 10^6$ Da) then the number of glutenin polymers (NP) in a wheat grain (of 40 mg weight, having 12% protein concentration, including 80% of storage proteins (50% gliadins and 50% glutenins), of which 25% are polymerised) was obtained as follows:

The mass of glutenin polymers would be $40 \times 0.12 \times 0.80 \times 0.5 \times 0.25 = 4.8 \times 10^{-4}$ g

Consequently, the number of glutenin polymers (NP) was:

$$NP = 6.02 \times 10^{23} \times 4.8 \times 10^{-4}/15 \times 10^6 = 1927 \times 10^9 \text{ polymers.}$$

Instead of using a constant distribution of the masses (Mw) in wheat grain, the number of glutenin polymers was computed for Gaussian distribution around the following average polymer masses and limit values, often found on classic measures using A4F.

| Average polymer mass avMw ($\times 10^6$ Da) | Limit values of the mass distribution | |
|---|---|---|
| | Min($-3\sigma$) | Max($+3\sigma$) |
| 3 | 0 | $6 \times 10^6$ |
| 8 | $5 \times 10^5$ | $15.5 \times 10^6$ |
| 15 | $6 \times 10^6$ | $24 \times 10^6$ |
| 30 | $15 \times 10^6$ | $45 \times 10^6$ |
| 60 | $30 \times 10^6$ | $90 \times 10^6$ |

For each of the average masses, the number of glutenin polymers was computed for grains having only the following kernel weights (30, 40, 50 and 60 mg) and the protein concentrations of 8.5, 10, 12, 14, 16 and 17%dw. The resulting 24 values of the total mass of glutenin polymers in wheat grain were used to compute the number of glutenin polymers based on a Gaussian distribution of the masses for each average polymer mass (Supplemental Figure S2 and Supplemental Table 2).

*Statistics*

As indicated in the first multi-location study, all statistical analyses, including Pearson correlations, analysis of variance (ANOVA) and partial least square regressions (PLS), were performed using Statgraphics® software. The broad-sense heritability, $H^2$, of the Conf parameter was computed using the GLM procedure of variance analysis, as previously described (Branlard *et al.*, 2020).

**Results and discussion**

*Polymer conformation exhibits a large phenotypic variation*

The conformation of the polymers had a global Gaussian distribution for the 885 samples analysed. However, an asymmetric profile was reported, showing an excess of values lower than the mean 0.496. The distribution evidenced a high range of phenotypic variations of the polymer conformation, from the lowest Conf value of 0.285, typical of a spherical polymer shape, to the highest value of 0.74, indicating a highly depolymerised macromolecule. These Conf values were significantly influenced by the 192 cultivars, as well as by the 11 locations in France and the 4 years of the wheat trials (2003–2004, 2004–2005, 2008–2009, and 2009–2010), as shown in Table 1. The percentages of the phenotypic variation explained by genotypes, locations and years were, however, lower for Conf compared to the percentages of its constitutive Mw and Rw parameters. The low broad-sense heritability coefficient ($H^2$ = 16.8) also confirmed the high influence of the environment (locations and years) on Conf values but may also have occurred because Conf is a computed ratio cumulating possible errors in measures on the two constitutive parameters.

**Table 1.** Values of statistical distribution (Mean, SD and Min-Max) of the 192 cultivars in 11 locations for the conformation of glutenin polymers (Conf). The percentage ($R^2$ %) of phenotypic variance was obtained using one-factor ANOVA, and the heritability ($H^2$) coefficient was averaged over $H^2$ computed per experimental year.

| Parameters (unit) | Total samples | Mean | SD | Min | Max | $R^2$ % Genotype (192) | $R^2$ % Location (11) | $R^2$ % Year (4) | $H^2$ (4) |
|---|---|---|---|---|---|---|---|---|---|
| Mw (KDa)* | 885 | 9554.4 | 5485.6 | 1142.0 | 48777.5 | 55.00 | 70.10 | 59.10 | 11.80 |
| Rw (nm)* | 885 | 42.40 | 20.94 | 6.70 | 116.20 | 79.6 | 85.20 | 82.60 | 25.00 |
| Mw/Mn (--)* | 885 | 12.58 | 11.81 | 1.04 | 82.94 | 61.50 | 85.60 | 74.10 | 21.20 |
| Conf (--) | 885 | 0.496 | 0.064 | 0.285 | 0.740 | 46.60 | 54.60 | 48.20 | 16.80 |

(*) Data from the multi-local trials previously reported by Branlard *et al.* (2020).

*Observation of the glutenin polymers using atomic force microscopy*

The lyophilised polymers, first observed using an optical microscope, showed numerous aggregates of polymers whose sizes were several microns (not shown in Figure 1). However, using AFM, the isolated GPs directly spread on silicon substrate from lyophilised glutenins were revealed for the first time. The AFM perfectly confirmed the A4F data: the height of isolated polymers is of nanometric size and, for the sample shown in Figure 1, rather of spherical shape. The profiles in Figure 1B revealed that isolated spheric polymers in contact of substrate, then a little spread out, are of less than 100 nm diameter; height and diameter polymer sizes revealed by AFM are in full agreement with A4F measures. Additional information on the GPs was shown by Cryo-TEM.

**Fig. 1.** Photos of the glutenin polymers issued from sequential glutenin extraction, as revealed using Cryo-Transmission Electron Microscopy**.** The black arrow points to a micelle (less than 10 nm in diameter). Short and long white arrows are explained in text.

*The glutenin polymer presents a micellar structure using Cryo-TEM*

The use of Cryo-TEM revealed a micellar structure of the GPs issued from sequential glutenin extraction (Figure 2). The micelles were not homogenously distributed in the vitrous ice film but form larger assemblies, such as a grape of micelles indicating that glutenins were aggregated. Four of these larger structures, of about 100 nm diameter, were encircled. In the lower left corner, an enlargement of one grape is shown. The short white arrow points towards a micelle probably corresponding to a glutenin subunit of less than 10 nm diameter and the long arrow points towards a strain of the polymer linking two micelles. These observations revealed that GPs, as evidenced in Figure 1, were formed of individual polymers (or micelles) mainly associated with hydrogen bonds but with possible covalent bindings. These observations were in full agreement with previous findings revealing that during grain desiccation, (1) the mass of the polymers drastically increased (Carceller and Aussenac, 1999) and (2) the abundance of the protein disulphide isomerases (PDIs) was significantly decreased (Branlard *et al.*, 2020). They also confirmed the importance of hydrogen bonding in gluten properties, as revealed by Belton (1999, 2005).

The microscopy techniques that were used here may provide new insights on studies focusing on the structure of the polymers. The spherical and more or less reticulated glutenin structure of macropolymers in 1.5% SDS solution was previously shown using Confocal Scanning Laser Microscopy technique (CSLM) (Don et al., 2003). Using for example near isogenic wheat lines, the Cryo-TEM together with AFM and CSLM will be useful to further analyse the influence of glutenin alleles on the structure of the polymers that were revealed different in mass, radius or conformation.

**Fig. 2.** Photos of the glutenin polymers issued from sequential glutenin extraction, as revealed using Cryo-Transmission Electron Microscopy**.** The black arrow points to a micelle (less than 10 nm in diameter). Short and long white arrows are explained in text.

GSPs, particularly glutenin polymers, are synthesised during compound accumulation of wheat grains in protein bodies whose biogenesis received much attention in cereals (for review in cereals see (Kumamaru *et al.*, 2007) and wheat (Shewry and Lafiandra, 2022). Protein bodies (PBs), which are derived from the endoplasmic reticulum (ER) inclusion membrane, were revealed to contain HMW-GS throughout grain development (Tosi et *al.*, 2009). Once released into the cytoplasm, the PBs had diameters varying between 0.2 and 2 μm. How many GPs can be stored in one PB? If we consider that all the polymers have a spherical shape and knowing their gyration radius (Rw), the hypothetical number of polymers stored in one PB can be estimated. If all GPs had an Rw of 5 nm (20 nm), their number would be as high as $10^6$ (more than 15 thousand) into one PB of 1 μm of diameter and $8 \times 10^6$ (125 thousand) into one PB of 2 μm diameter (See Table S1).

*Simulation of the number of glutenin polymers (GPs) in wheat grain and flour*

Numerous factors are involved in the amount of polymerised glutenin in wheat grain. Together with the average mass of polymers (avMw), two major factors, kernel weight (KW) and grain protein concentration (PC), were used to compute the number of polymers in the wheat grain. One hundred and twenty combinations (4 KW, x 6 PC x 5 avMw) were used for an amount of polymerised glutenin per grain between 0.255 and 1.02 mg. Considering the different avMw, with individual masses following a Gaussian distribution, the number of GPs varied between $2.55 \times 10^{12}$ and $204.68 \times 10^{12}$ per grain (Table S2). The number of polymers per grain for a given protein concentration was the highest (lowest) when polymers were of low (high) average mass for bigger (smaller) grain weight (Figure 3A). Similarly, the number of polymers per grain for a given grain weight was higher (lower) with high (low) PC values. Table S2 also shows the number of GPs for 100 gr of flour with 6 PCs. The amount of polymerised glutenin for PCs varying between 8.5 and 17%dw was between 0.85 and 1.7 g in these flours, giving a range of the expected number of polymers between $8.528 \times 10^{15}$ and $341.13 \times 10^{17}$ for 60 and 3 MDa of average polymer mass, respectively (Figure 3B). The calculated polymer numbers were obtained for an average percentage of 25% polymers in the glutenin fraction. Many genetic and environmental factors can influence the percentage of the polymerized fraction of glutenins and obviously may cause variation in the above calculated numbers as previously reported (Branlard et al., 2020), genetic and environmental factors are significantly associated with variations in polymer mass. The following sections aim to test the possible factors influencing the polymer conformation.

**Fig. 3A.** Glutenin polymer number ($\times 10^{12}$) per grain as a function of average polymer mass and kernel weight for grain having a protein concentration of 12% dw.

**Fig. 3B.** Glutenin polymers number (×10$^{15}$) in 100 g of flour as a function of average polymer mass and grain protein concentration.

*Allelic variants of glutenin have a limited influence on polymer conformation*

The protein concentration and grain hardness have very low but significant influences on the phenotypic variations of Conf. These two variations in grain quality characteristics were reported as significantly decreasing the polymer mass (Mw), whereas the gyration radius (Rw) was positively increased by PC and decreased by GH (Figure S3; Branlard *et al.*, 2020). The hard type of wheat kernel has been shown to be associated with the unfolded protein response (UPR), inducing earlier endosperm cell death (Lesage *et al.*, 2012). This UPR probably resulting from oxidative stress caused earlier shutdown of polymer accumulation and lower values for polymer mass and gyration radius. The Conf was the coefficient obtained from the slope (1/a) of the linear relationship of Log (Rw) = f [Log (Mw)]. Conf variations were then logically associated with Rw, decreasing with PC and increasing with GH (Figure 4). As reported in Table 2, PC and GH together explained only 1.1% of the phenotypic variations in the Conf values, as revealed by the PLS regression involving the 885 samples analysed from 11 locations. When the 14 HMW-GS alleles were introduced in the PLS regression, in addition to PC and GH, explanations of the Conf variations were also rather low (4.66%). When the 17 LMW-GS alleles were introduced instead of the HMW-GS alleles, the percentage explained of phenotypic Conf variations increased to 10.75%. A previous study reported that LMW-GS diversity contributed two times more than HMW-GS for Mw and Rw variations; here, LMW-GS explained more than two times the variation explained by HMW-GS in Conf variations. The total contribution of the glutenin alleles (together with PC and GH) was 14.21% (Table 2) lower than the percentage reported (Branlard *et al.*, 2020) for its components, Mw and Rw (20.78% and 25.49%, respectively). The specific contributions of the alleles encoding HMW-GS and LMW-GS in the Conf values resulting from the PLS regressions using the 885 wheat samples are shown in Figure 4.

**Table 2.** Contribution (%) of the total phenotypic variance of the polymer conformation (Conf) explained by the following variables using partial least square regression (PLS): grain protein concentration (PC), grain hardness (GH), high molecular weight glutenin subunits (HMW-GS) alleles, low molecular weight glutenin subunits (LMW-GS) alleles, cumulative mean temperature (Tmean) and cumulative water precipitation (WatSum) in the three final months in the 11 experimental fields of wheat crops involving 192 cultivars and previously detailed data (Branlard *et al.*, 2020). With: n= 885; (**) p<10$^{-2}$ ; (***) p<10$^{-4}$.

| PC + GH | PC + GH + HMW-GS | PC + GH + LMW-GS | PC + GH + HMW-GS + LMW-GS | PC + GH + Tmean May + Tmean Jun + Tmean Jly | PC + GH + Tmean May + Tmean Jun + Tmean Jly + HMW-GS + LMW-GS | PC + GH + WatSum May + WatSum Jun + WatSum Jly + Tmean May + Tmean Jun + Tmean Jly + HMW-GS + LMW-GS |
|---|---|---|---|---|---|---|
| 1.11** | 4.66*** | 10.75*** | 14.21*** | 41.34*** | 42.37*** | 58.64*** |

The figure clearly shows the high positive coefficients of all *Glu-B1* alleles contributing to an increase in the Conf values. All alleles encoded at *Glu-B1* decreased both parameters, Mw and Rw (Branlard *et al.*, 2020; Figure S3). These results clearly confirm the hypothesis that was proposed, using a proteomic approach, about the involvement of asparaginyl endopeptidases in the post-translational processing of the By-type subunits (Nunes-Miranda *et al.*, 2017); these asparaginyl endopeptidases are cysteine endopeptidases also described to mediate the maturation process of other storage proteins, such as 2S albumins and 11S globulins (Hara-Nishimura *et al.*, 1995; Tan-Wilson and Wilson, 2012). The proteolytic cleavage observed for the By-type glutenin subunits leads to the loss of the unique conserved cysteine residue in the C-terminal domain. The resulting reduced polymerisation was more pronounced on the gyration radius Rw than on the polymer mass (Mw); consequently, the Conf parameter increased (Figure 4). In addition to the seven *Gli-B1* alleles (namely 6-8, 7, 7-8, 7-9, 13-16, 14-15, 17-18) the following five LMW-GS alleles had a greater positive contribution, increasing the Conf value: *Glu-A3a* and *b*, *Glu-B3a* and *b*, and *Glu-D3a*. Moreover, the well-known *Glu-D1d* allele, subunits 5-10, slightly increased the conformation of the polymers, whereas null alleles, such as *Glu-A1c*, *Glu-A3e* (here noted *Glu-A3ef*) and *Glu-B3j* (translocation 1BL-1Rs), decreased the Conf value. Interestingly, strongly decreased Conf values were found for alleles that were previously ranked between medium to a rather good effect on rheological properties (Branlard et al., 2001), such as *Glu-D1b* 3-12, *Glu-B3g* and *Glu-D3b*. Their subunits increased Mw and Rw (Figure S3; Branlard *et al.*, 2020), but the polymerisation process had a greater impact on the gyration radius (Rw) than on Mw. Although the glutenin alleles had limited influence on the Conf values, they may be considered for genetic breeding aimed at improving the degradability of glutenin polymers, which is obviously expected for a Conf higher than 0.5 (corresponding to depolymerised polymers).

**Fig. 4.** Standardised coefficients of the PLS regressions aimed at explaining the conformation (Conf) values of the glutenin polymers measured using A4F-MALLS over the 885 wheat samples. The protein concentration of the grain (PC; noted protein) and grain hardness (GH; noted hardness) were explanatory variables first introduced in the regression, followed by alleles of HMW-GS and LMW-GS.

*High temperatures during grain maturation increase the compactness of glutenin polymers*

Considering the variables PC, GH and 31 alleles encoding the glutenin subunits, the cumulative daily average temperature of the last three months (TmeanMay, TmeanJune and TmeanJuly) of the wheat crop were introduced as three additional explanatory variables in the PLS regression. As was the case for the previous variables (Mw, Rw and PI), the explanation of the phenotypic Conf values was greatly increased (from 14.21 to 42.37%, Table 2) by the cumulative daily average temperature occurring in these three months in the 11 experimental locations. The standardised coefficients attributed to all variables clearly showed the greater influence of the temperature of the last month, July, which corresponds to grain maturation (Figure S3a, S3b and S3c). The atmospheric temperature during June increased the Conf value in allowing normal storage protein accumulation, increasing Mw and Rw without any physiological perturbations. In June, the photosynthetic carbohydrates and nitrogen remobilisation were actively engaged in grain filling, generally without severe heat and drought, compared to less favourable conditions that occurred during July in France. Considering June, the average daily temperatures (day + night) were, from location to location, between 15.7 and 18.7°C for the mean temperature and between 21.3 and 24.6°C for the maximum registered temperature, whereas for July, these values were 17.2–21.8°C and 23.7–27.9°C, respectively. The simple regression of the Conf values on the cumulative daily mean temperature measured in 11 locations during July clearly revealed that lower Conf values were caused by higher temperatures (Figure 5). The Conf values of the glutenin polymers were greatly reduced by the high temperatures occurring in the last month of the wheat cycle. The partially depolymerised and reticular polymers usually observed for a Conf value around 0.5 acquired a compact and spherical shape (Conf around 0.35) under the effect of higher temperatures during grain maturation.

**Fig. 5.** Regression between the conformation of the glutenin polymers, measured using asymmetric flow field-flow fractionation multi-angle laser light scattering (A4F-MALLS), and the cumulative daily mean temperatures (°C) in July (noted TmeanJly) in the 11 wheat trial locations in France between 2004 and 2010, involving 192 hexaploid cultivars.

By analysing the glutenin polymer variations using A4F-MALLS, we revealed that the daily average temperature (i.e., 3.5°C of difference per day between the two extreme thermal locations during the last month of grain development, caused an increase of more than 189% and 242% in Rw and Mw, respectively (Branlard et al., 2020). As opposed to the two above polymer parameters the Conf parameter was instead decreased when the daily average temperature increased. According to the regression equation (Figure 5), the mean Conf values decreased by 20.79% between the two extreme thermal locations. This revealed a major influence of the daily average temperature on the gyration radius of the polymers (Rw) compared to the mass of the polymers (Mw). In addition, the simple Pearson correlation of phenotypic values between Conf and Mw or Rw, computed in each of the 11 locations, revealed no consistent correlations (see Table S3). These results indicate that polymerisation involves a more complex process than the simple addition of glutenin subunits, increasing both Mw and Rw.

*Limits of the redox system in hypoxic environments*

The phenotypic Conf values revealed noticeable variations within locations, particularly due to genotypic differences, maturity earliness and resistance to water stress. Rainfall was also registered over the last three months in each trial location. The average rainfall in May, June and July significantly influenced Mw and Rw (Branlard *et al.*, 2020). The percentage of explained phenotypic Conf values over the 11 locations, including the 885 wheat samples, increased from 42.34 to 58.67% (Table 2). In particular, July rainfall contributed to an increase in Rw, which, together with the cumulative daily average temperature of that month, decreased the Conf values (Figure S3c). PC was the grain characteristic with the highest significant influence on reducing the Conf value of the glutenin polymers (Figure S3c). The PLS regression clearly showed that the coefficient attributed to PC was more important than any other coefficient attributed to glutenin alleles or to July rainfall in reducing the polymer Conf value. This association of PC in the explanation of the reduction of the Conf value can be seen as a "marker" of environmental factors occurring during polymerisation. Higher temperatures and heat stress reduce photosynthesis, photo-assimilates and starch synthesis in grain, thereby increasing the PC.

The water uptake in July may have partially alleviated the oxidative stress occurring in wheat endosperm during that month, since the majority of nitrogen translocation in the spike and protein accumulation in grain is usually achieved at the beginning of July in France. The polymerisation peak was shown to occur at the end of the water plateau, that is, the initiation of dehydration and grain maturation (Carceller and Aussenac, 1999; Koga *et al.*, 2017). The process of storage protein accumulation and polymerisation in ER involves a tremendous number of SS links catalysed by oxidative molecules, such as protein disulphide isomerases (PDI), sulfhydryl oxidases and reactive oxygen species (ROS), such as hydrogen peroxide ($H_2O_2$), hydroxyl radical ($OH^-$) and super oxide ($O2^{--}$; Sevier and Kaiser 2008, Lal *et al.*, 2021). Plants may reduce the excess and detrimental effects of ROS with several antioxidant molecules, such as ascorbic acid (ASC) and glutathione (GSH), and scavenging enzymes, such as ascorbate peroxidases (APX), glutathione reductases (GR), superoxide dismutases (SOD) and catalases (CAT). Proteomic analysis of the developing endosperm of wheat grown in normal controlled conditions offered the opportunity to identify the expression of the 19 following enzymes involved in ROS scavenging: 7 APX, 3 Dihydro-ascorbate reductases (DHAR), 5 SOD, 1 Glutathione S-transferase (GST), 1 Thioredoxin reductase and 1 1-Cys-peroxyredoxin. The evolution of these ROS scavenging enzymes (Figure 6), expressed in the cytosol of the endosperm, during the 21 sampling stages (Tasleem-Tahir *et al.*, 2012), clearly showed the decreasing abundance for APX, CAT and DHAR and increasing abundance for GST, NADPH thioredoxin reductase and SOD of wheat grain after 500°CDay and maturation.

**Fig. 6.** Evolution of the major enzymes involved in the Redox function, revealed using proteomics analysis over 21 stages of grain formation and maturation (Tasleem-Tahir *et al.*, 2012). Evolution of the average of normalised spot volumes for 7 APX, 3 DHAR, 5 SOD, 1 Catalase, 1 GST, 1 NADPH Thioredoxin reductase and 1 1-Cys Peroxiredoxin reductase during grain formation, in cumulative thermal time (°Cd). Red continuous and blue dotted lines indicate increased and decreased abundance, respectively, for enzymes during polymer formation and grain desiccation; for each individual enzyme, the confidence interval (not shown for clarity of the figure) of each data point was between 5 and 12% of the data mean value computed on 6 replicates of 2DE.

The controlled conditions for the wheat growing environment were chosen as normal conditions with a full nutritional and water supply without any stress. These enzymes were reported to be overexpressed in response to heat stress during grain compound accumulation (Skylas *et al.*, 2002; Majoul *et al.*, 2003, 2004, 2013; Vensel *et al.*, 2005; Hurkman *et al.*, 2009; Laino *et al.*, 2010). This overexpression may be insufficient to properly oxidise the free thiols potentially engaged in the polymerisation process in the small protein bodies (PBs) sequestered in the ER (Galili *et al.*, 1996). The post-translational processing of the Glu-B1y subunits, as evidenced in the 192 wheat varieties, removed their cysteine residue in the C terminal domain and consequently had limited branching in subunits. The excess ROS causes the unfolded protein response (UPR) initially described in yeast and mammals cells and has been shown to induce numerous signal transduction pathways involving many genes, impacting various processes, such as lipid biosynthesis, N-linked glycosylation, ER-associated degradation (ERAD), vacuolar function and chaperones (Jonikas *et al.*, 2009). The UPR was revealed in different plant tissues, including the wheat endosperm, in relation to grain hardness (Lesage *et al.*, 2012) and can be elicited during plant development in regulating basal immunity responses (Bao *et al.*, 2017). Because of excess ROS, the phospholipids forming the lipidic bilayer polar membrane of ER are degraded. It can be hypothesised that this degradation of membrane lipids by hydroperoxide prevents the ER from performing additional membrane inclusions to surround more protein bodies. The accumulation of polymers thus continues in the PBs already present in the ER. The polymers are then packed in fewer PBs, as they should be, and reach higher mass, becoming less reticulated for less volume hindrance and obtaining a global spheric shape. The insufficient amount of phospholipids forming the cytoplasm membranes, such as for ER and PBs, is strongly emphasised in wheat under heat stress. A survey of 539 wheat accessions, cultivated in France during the two last centuries, revealed a significant opposition between the oldest and modern cultivars; the former being richer in poly-unsaturated fatty acids and the latter with more saturated fatty acid (stearic, palmitic) and mono-unsaturated oleic acid (Roussel *et al.*, 2005). Modern cultivars had more phospholipids, but this seemed insufficient to allow them to respond to the current heat stress. The use of marker-assisted selection in breeding wheat with higher polymer conformation, together with agronomic practices aimed at improving soil water retention, should help to reduce the consequences of high temperatures (Chaudhary *et al.*, 2020; Lal *et al.*, 2021), particularly for wheat glutenin polymers.

**Conclusions**

As noted in a previous companion publication (Branlard *et al.*, 2020), the high temperatures registered in France during our four experimental years can also be encountered in many other countries. The influence of high temperatures is the cause of higher masses and a higher gyration radius of polymers. Importantly, we revealed here that the polymer becomes less reticulated, more spherical, and more compact when the average temperature of the last month of the wheat cycle is increased. These observations make the very numerous glutenin polymers the most reliable hypothesis to explain non-celiac gluten sensitivity (NCGS) that many people suffer from in countries subjected to global warming. The polymers of glutenin have to be seized by wheat geneticists, technologists, food manufacturers and nutritionists for developing healthier foods for consumers.

There are numerous fields to investigate. Basic research on molecular mechanisms occurring in endoplasmic reticulum involved in the polymerisation process and in redox system or oxidative stress has to be investigated specifically with wheat grown in controlled environments. To face the global warming consequences, geneticists can use glutenin alleles, increasing the polymer conformation and reticulation. Numerous allelic effects remain to be tested from world wheat collections and several genetic approaches are available today for breeding future wheat possessing better control of the characteristics of glutenin polymers accumulated in high temperature environments. These investigations will take years and we fear that environmental factors may lead to variations in polymer characteristics in the years to come. The technologists, more rapidly, may develop adapted processes with the aim of manufacturing healthy, baked products for consumers. During bread baking, the amylose released from the starch granules can surround the gluten films, yielding gelatinised amylose; this resistant starch prevents the rapid hydrolysis of these glutenin polymers, which are more compact and less reticulated today. The new processes should obviously include longer fermentation times compared to those routinely practiced today and a controlled hydrolysis of glutenin polymers. Nutritionists are also encouraged to measure polymer characteristics that could be digestible without any trouble in the intestines of sensitive people. The above genetic and environmental information on the polymer characteristics of wheat should help the scientific community to propose an efficient answer for people suffering from NCGS. Moreover, considering that Alpha-, Beta-, and Gamma-gliadin bound by an additional odd SH to glutenins can be sequestered in these polymers, the present findings could be of some help to alleviate other pathologies linked to gluten consumption.

**Supplementary data**

The following supplementary data are available at JXB online.

**Table S1.** The simulated number of glutenin polymers accumulated in protein body having either 1- or 2-micron diameter is given.

**Table S2.** The glutenin polymer number in wheat grain and flour is simulated assuming gaussian distribution of the polymer masses.

**Table S3.** Pearson correlation between Phenotypic values of glutenin polymers characteristics (Conf-Mass Mw) and (Conf- Radius Rw) of the polymers as measured using A4F-MALLS.

**Fig. S1**. The Plot of the linear relationship between Log (Rw) and Log (Mw) of glutenin polymers extracted for a wheat sample is shown.

**Fig. S2**. The Gaussian distribution of the number of polymers in wheat grain with given characteristics is exemplified.

**Fig. S3a.** Standardized coefficients of the PLS regressions aimed to explain the mass MW of the polymers.

**Fig. S3b.** Standardized coefficients of the PLS regressions aimed to explain the gyration radius Rw of the polymers.

**Fig. S3c.** Standardized coefficients of the PLS regressions aimed to explain the Conf of the polymers.

**Acknowledgements**

The authors thank the FSOV (Fond de Soutien à l'Obtention Végétale) for its financial support (programme "Indice of Quality" and programme "Tenacity—Extensibility Bread Loaf Volume"—Leader G. Branlard, INRAE, UCA UMR1095 GDEC) and the majority of European private breeding companies operating in France (Caussades semences, Desprez, Momont-Hennette, Saaten Union, Serasem, Syngenta Seeds, and Verneuil LVH) for the provision of their wheat cultivars and their logistical support. The authors also thank all engineers and technicians of the private breeding companies and public institutes who participated in the FSOV—particularly, Sylvie Dutriez (Caussades semences), Jane Stragliati (Verneuil LVH), Volker Lein (Saaten Union), Philippe Lonnet et Thierry Demarquet (F Desprez), Eric Margale (Serasem), Benoit Méléard (ARVALIS), Fabrice Lagoutte and François Xavier Oury (INRAE), Patrice Senellart (Syngenta Seeds), and Stephen Sunderwirth (Momont-Hennette). Catherine Ravel (INRAE Clermont Ferrand) and Didier Marion (INRAE Nantes) are greatly acknowledged for their useful discussions. Julien Branlard is also greatly thanked for his help with Gaussian simulation.

**Author contributions**

All authors contributed significantly to the research. G.B. designed the experimental studies in collaboration with T.A. and FSOV breeding members. INRAE Clermont Ferrand and FSOV breeding companies performed the experimental trials. A.D. and M.S. performed AFM and Cryo-TEM, respectively. Proteomics analyses were carried out by A.T. A.F. identified glutenin alleles. L.R. and T.A. performed A4F-MALLS characterisation of the glutenin polymers. G.B. performed the statistical analyses and Gaussian simulations. G.B., T.A., L.R., A.D., M.S, A.T. and A.F. contributed to the writing of the final manuscript. All authors have read and agreed to the published version of the manuscript.

**Conflicts of Interest**

The authors declare no conflicts of interest. The funders had no role in the design of the study; in the collection, analyses or interpretation of data; in the writing of the manuscript; or in the decision to publish the results.

**Data availability**

The data supporting the findings of this study are available from the corresponding authors, (Gérard Branlard and Thierry Aussenac), upon request.

**Funding**

This research was partly funded by FSOV (Fond de Soutien à l'Obtention Végétale) and FSOV breeding members, programme "Indice of Quality" and programme "Tenacity—Extensibility Bread Loaf Volume."

**Tables legends**

**Table 1.** Values of statistical distribution (Mean, SD and Min-Max) of the 192 cultivars experimented in 11 locations for the Conformation of glutenin polymers (Conf). The percentage ($R^2$ %) of phenotypic variance was obtained using one factor ANOVA, and the heritability $H^2$ coefficient was averaged over the $H^2$ computed per experimental year.

**Table 2.** Part (%) of the total phenotypic variance of the polymers conformation (Conf) explained by the following variates using partial least square regression (PLS): grain Protein Concentration (PC), Grain Hardness (GH), High Molecular Weight Glutenin Subunits (HMW-GS) alleles, Low Molecular Weight Glutenin Subunits (LMW-GS) alleles, cumulative mean temperatures (Tmean) and cumulative water precipitations (WatSum) for the three final months in the 11 experimental fields of wheat crop involving 192 cultivars and previously detailed [Branlard *et al.*, 2020].

**Figures legends**

**Fig. 1**. A) AFM error image of lyophilised glutenin polymers on a silicon substrate; B) the topographic image corresponding to the error image in A) and; C) polymer profiles. The areas where the height profiles were measured are indicated with green lines on image B).

**Fig. 2.** Photos of the glutenin polymers issued from sequential glutenin extraction, as revealed using Cryo-Transmission Electron Microscopy. The black arrow points to a micelle (less than 10 nm in diameter). Short and long white arrows are explained in text.

**Fig. 3A.** Glutenin polymer number ($\times 10^{12}$) per grain as a function of average polymer mass and kernel weight for grain having a protein concentration of 12% dw.

**Fig. 3B.** Glutenin polymers number ($\times 10^{15}$) in 100 g of flour as a function of average polymer mass and grain protein concentration.

**Fig. 4.** Standardised coefficients of the PLS regressions aimed at explaining the conformation (Conf) values of the glutenin polymers measured using A4F-MALLS over the 885 wheat samples. The protein concentration of the grain (PC; noted protein) and grain hardness (GH; noted hardness) were explanatory variables first introduced in the regression, followed by alleles of HMW-GS and LMW-GS.

**Fig. 5**. Regression between the conformation of the glutenin polymers, measured using asymmetric flow field-flow fractionation multi-angle laser light scattering (A4F-MALLS), and the cumulative daily

mean temperatures (°C) in July (noted TmeanJly) in the 11 wheat trial locations in France between 2004 and 2010, involving 192 hexaploid cultivars.

**Fig. 6.** Evolution of the major enzymes involved in the Redox function, revealed using proteomics analysis over 21 stages of grain formation and maturation (Tasleem-Tahir *et al.*, 2012). Evolution of the average of normalised spot volumes for 7 APX, 3 DHAR, 5 SOD, 1 Catalase, 1 GST, 1 NADPH Thioredoxin reductase and 1 1-Cys Peroxiredoxin reductase during grain formation, in cumulative thermal time (°Cd). Red continuous and blue dotted lines indicate increased and decreased abundance, respectively, for enzymes during polymer formation and grain desiccation; for each individual enzyme, the confidence interval (not shown for clarity of the figure) of each data point was between 5 and 12% of the data mean value computed on 6 replicates of 2DE.

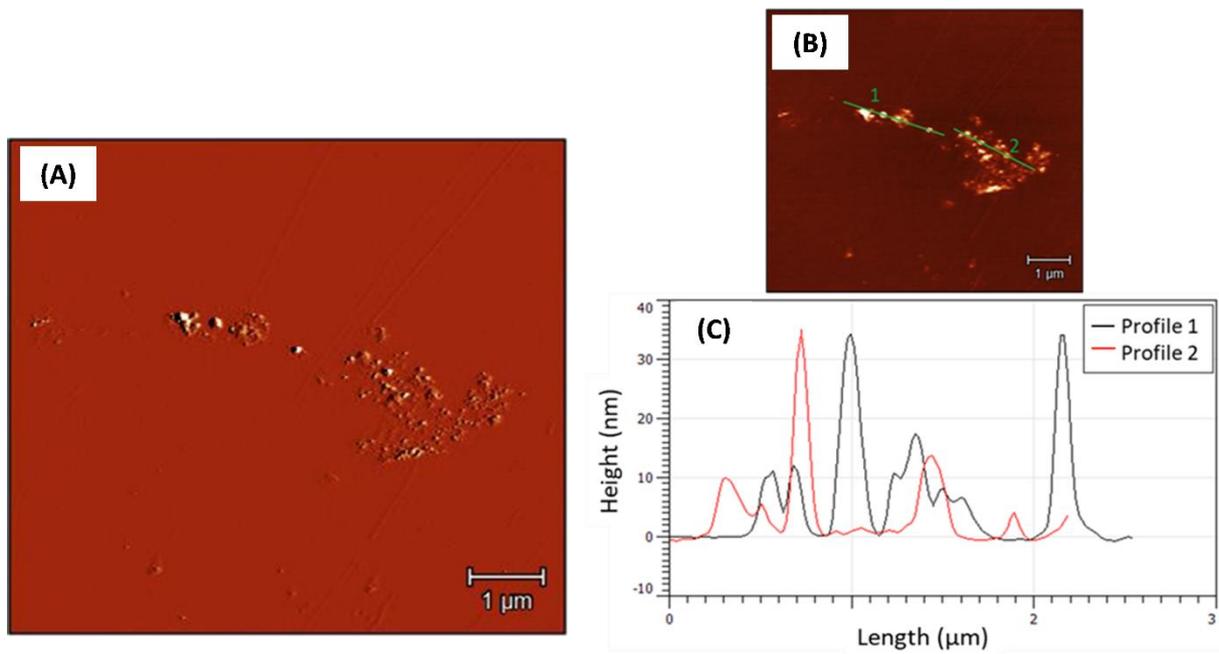

**Fig. 1**

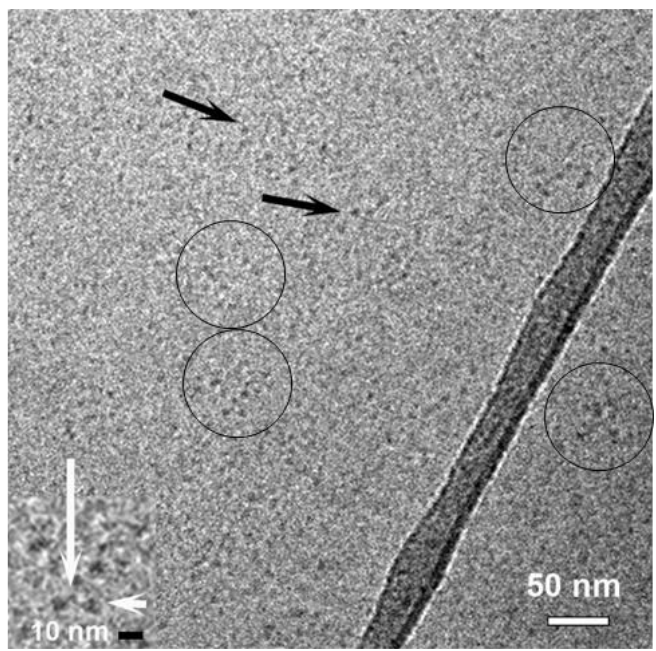

**Fig. 2**

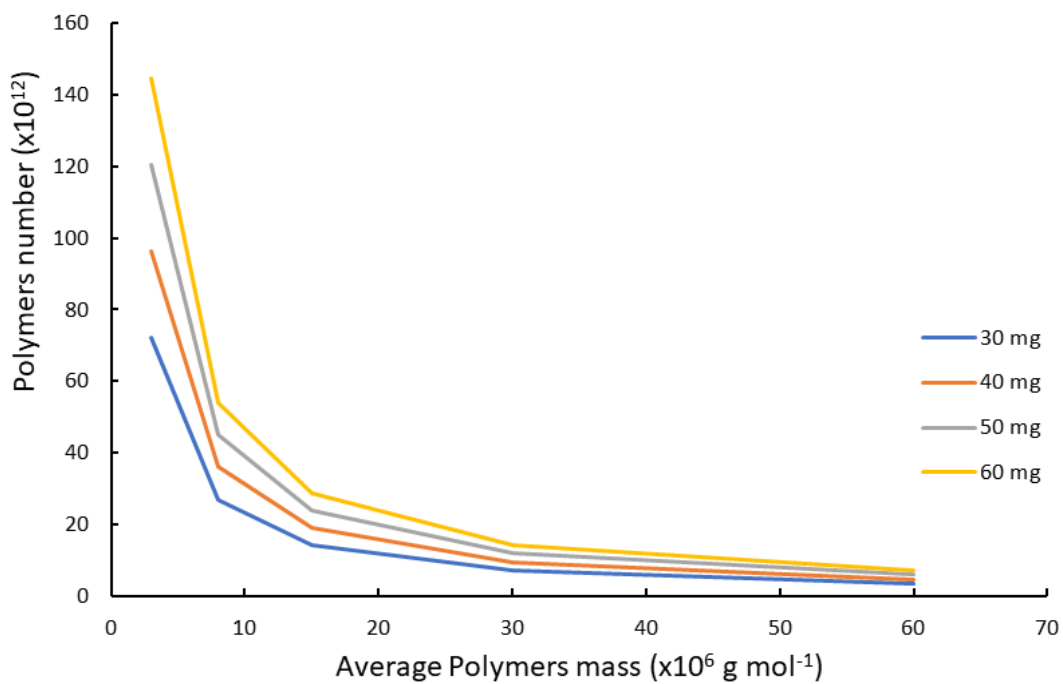

**Fig. 3A**

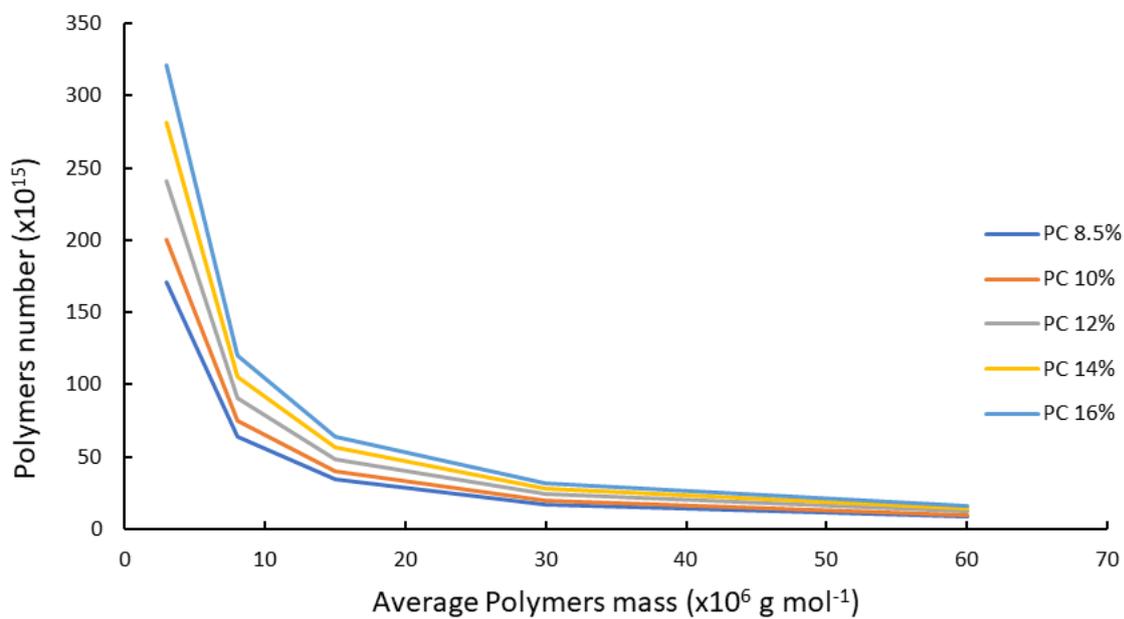

**Fig. 3B**

**Fig. 4**

**Fig. 5**

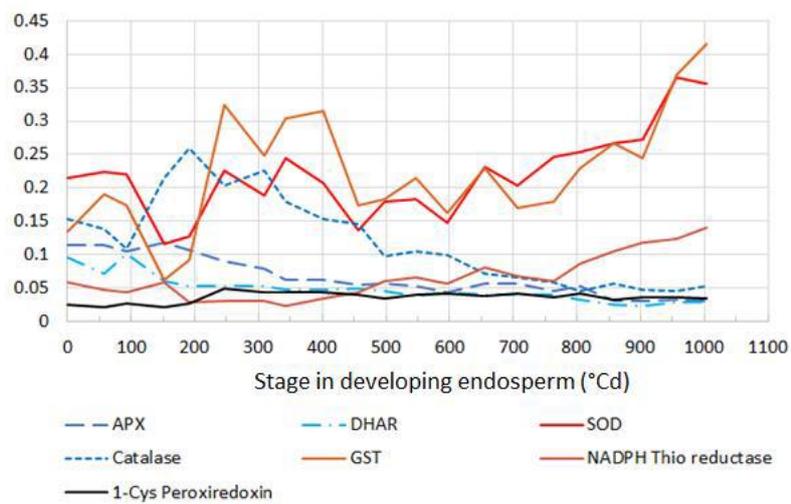

**Fig. 6**